\documentclass[nofootinbib,showpacs,pra,onecolumn,aps,floatfix]{revtex4}
\usepackage{color}
\usepackage{graphicx}
\begin{document}
\title{Diffraction in time of a confined particle and its Bohmian paths}
\author{S. V. Mousavi}
\email{vmoosavi@qom.qc.ir}
\affiliation{Department of Physics, The University of Qom, P. O. Box 37165, Qom, Iran}
\begin{abstract}
Diffraction in time of a particle confined in a box which its walls are removed suddenly at $t=0$ is studied. The solution of the time-dependent Schr\"{o}dinger equation is discussed analytically and numerically for various initial wavefunctions. In each case Bohmian trajectories of the particles are computed and also the mean arrival time at a given location is studied as a function of the initial state.
\end{abstract}
\maketitle
\section{introduction}
Several works have been done on time-dependent boundary conditions \cite{Mo-PR-1952,GeKa-Sov-1976,GaGo-Zphys-1984,BaMaHo-PRA-2002,ManMajHo-PLA-2002,MaHo-Pra-2002}. Diffraction in time was initially 
introduced by Moshinsky \cite{Mo-PR-1952}. A beam of particles impinging from the left on a totally
absorbing shutter located at the origin which is suddenly turned off at time zero. The transient current
has a close mathematical resemblance with the intensity of light in the Fresnel diffraction by a straight edge.
An interesting feature of the solutions for cutoff initial waves, occurring both in the free case \cite{Ho-book-1993} and in the presence
of a potential interaction \cite{CaRu-PRA-1997}, is that, if initially there is a zero probability for the particle to be at $x > 0$, as soon as $t = 0^+$, there is instantaneously, a finite, though very small, probability to find the particle at any point $x > 0$. 
This non-local behavior of the Schrödinger solution is due to its nonrelativistic nature and not a result of the quantum shutter setup \cite{GaCaVi-PRA-1999}.
Application of the Klein-Gordon equation to the shutter problem \cite{Mo-PR-1952} shows that the probability density is restricted to the accessible region $x < ct$ ($c$ is the speed of light). 
See \cite{CaGrMu-PR-2009} for a recent review. Gerasimov and Kazarnovskii \cite{GeKa-Sov-1976} confined the initial wave in a finite region by introducing a second shutter at the point $x=L$. Godoy \cite{Go-PRA-2002} pointed out the analogy with Fraunhoffer diffraction in the case of small box (compared to the de Broglie length), and Fresnel diffraction, for larger confinements. 
In this context, by considering the problem of a particle in a one-dimensional box potential which its walls are suddenly removed at some time, the aim of the present paper is to probe some aspects of the time-dependent boundary condition for a particle confined in an square well focusing on Bohmian interpretation of quantum mechanics that have remained hitherto unnoticed. 
The computed Bohmian trajectories are instructive in revealing the conceptual ramifications of such an example. 

Although the formalism of Bohmian mechanics does not give predictions going beyond those of QM whenever the predictions of the later are unambiguous, it should be favored because of its interpretational advantages stemming from the ontological continuity between the classical and the quantum domains \cite{Ma-FoundPhys-2009}. Noticing the Bohmian arrival time formulation by means of cut-off current, it has been argued predictions of Bohmain mechanics are in contradiction to the standard quantum mechanical formalism \cite{RuGrKr-JPA-2005}.
In nonrelativistic Bohmian mechanics the world is described by point-like particles which follow trajectories determined by a law of motion. The evolution of the positions of these particles are guided by a wavefunction which itself evolves according to the Schr\"odinger equation \cite{BoI-PR-1952, BoII-PR-1952, HiBo-book-1993, Ho-book-1993, DuGoZa-1992-JSP}. In this theory, in the absence of any measuring device, one finds \cite{Le-PhyslettA-1993,Le-book-1996,Le-PRA-1998} that for those particles that actually reach $x=X$, the arrival-time distribution is given by the modulus of the probability current density, i.e., $|j(X, t)|$.
We will proceed as follows. In Sec. \ref{Sec: frprop} solution of time-dependent Schr\"{o}dinger equation is given for a particle which is initially confined in a box. Sec. \ref{Sec: BoTr} contains a very brief review of relevant parts of Bohm's interpretation of quantum mechanics. Sec. \ref{Sec: NuCa} gives numerical results. Finally, in Sec. \ref{CoRe} we present the concluding remarks.
\section{Free propagation of a particle initially confined in an square well} \label{Sec: frprop}
Consider a particle which is initially confined in an interval $[0, L]$ with wavefunction $\psi_0(x)$. If at time $t=0$ it is free, then at any instant $t>0$ its wavefunction is given by, 
\begin{eqnarray}
\psi(x, t) &=& \int_{-\infty}^{\infty} G(x, t|x^{\prime}, 0)\psi_0(x^{\prime})~dx^{\prime},
\end{eqnarray}
in which $G(x, t|x^{\prime}, 0)$ is the free particle propagator and is determined by,
\begin{eqnarray}
G(x, t|x^{\prime}, 0) &=& \sqrt{\frac{m}{2\pi i \hbar t}} e^{\frac{im}{2\hbar t} (x-x^{\prime})^2}~,
\end{eqnarray}
and $\psi_0(x^{\prime})$ is the initial wavefunction.
In this work we take initial wavefunction to be, (a) a stationary state of a particle inside a well with hard (perfect reflective) walls at $x=0$ and $x=L$ and (b) a motionless localized Gaussian wave-packet in that region with negligible overlap with the walls of the well. To avoid any problem concerning the boundary conditions, one can suppose in this case that the walls act as absorbers or the tails of the wave packet has been cut by the walls of the well.
In the first case initial wavefunction is given by, $\psi_0(x)=\phi_n(x)=\sqrt{2/L}\sin({k_n x})~\chi_{[0, L]}(x)$, with $k_n=n\pi/L$. $\chi_{[0, L]}(x)=\Theta(L-x)-\Theta(-x)$ is the characteristic function in the interval $[0, L]$. Simultaneous removal of both walls leads at time $t$ to the wavefunction \cite{Ho-book-1993,Go-PRA-2002,CaMu-EPL-2006}
\begin{eqnarray} \label{eq: wf_n}
\psi _n(x, t) &=& \sqrt{\frac{m}{2\pi i \hbar t}} \sqrt{\frac{2}{L}} \int_0^L e^{\frac{im}{2\hbar t} (x-x^{\prime})^2} \sin({k_n x^{\prime}})~dx^{\prime}~, \nonumber\\
&=& 
\sqrt{\frac{m}{4\pi i^3 \hbar t L}} \int_0^L e^{\frac{im}{2\hbar t} (x-x^{\prime})^2} \left( e^{ik_n x^{\prime}} -e^{-ik_n x^{\prime}}\right) 
~dx^{\prime}~, \nonumber\\
&\equiv&
\psi _{n,+}(x, t) + \psi _{n,-}(x, t)~,
\end{eqnarray}
which is a superposition of a right and a left moving diffracted in time plane waves. After doing some simple algebra, one gets
\begin{eqnarray}
\psi _{n,+}(x, t) &=& \frac{1}{\sqrt{4i^3L}} e^{ik_nx-iE_n t/\hbar} [F_n(x-L, t)-F_n(x,t)]~,\\
\psi _{n,-}(x, t) &=& \frac{1}{\sqrt{4i^3L}} e^{-ik_nx-iE_n t/\hbar} [F_n(L-x, t)-F_n(-x,t)]~,
\end{eqnarray}
with $E_n=\hbar^2 k_n^2/2m$ and 
\begin{eqnarray}
F_n(x, t) &=& \int_0^{\xi_n (x, t)} du e^{i\pi u^2/2}~,
\end{eqnarray}
with upper limit $\xi_n (x, t)= \sqrt{\frac{m}{\pi \hbar t}} \left(v_n t-x \right)$ in which $v_n=\hbar k_n/m$. 
Let us for later use, compute the derivative of $\psi _n(x, t)$ with respect to $x$.
\begin{eqnarray}
\frac{\partial \psi _{n,+}(x, t)}{\partial x} &=& ik_n \psi _{n,+}(x, t) + e^{ik_nx-iE_n t/\hbar} [\frac{\partial F_n(x-L, t)}{\partial x}-\frac{\partial F_n(x,t)}{\partial x}]~,\\
\frac{\partial \psi _{n,-}(x, t)}{\partial x} &=& -ik_n \psi _{n,-}(x, t) + e^{-ik_nx-iE_n t/\hbar} [\frac{\partial F_n(L-x, t)}{\partial x}-\frac{\partial F_n(-x,t)}{\partial x}]~,
\end{eqnarray}
in which,
\begin{eqnarray}
\frac{\partial F_n(x, t)}{\partial x} &=& -\sqrt{\frac{m}{\pi \hbar t}} e^{\frac{i \pi \xi_n ^2(x, t)}{2}}~.
\end{eqnarray}
Now, by some straightforward algebra, one can show,
\begin{eqnarray}\label{eq: wf_der}
\frac{\partial \psi_n}{\partial x} \big|_{x=L/2} &=& 2 e^{-iE_n t/\hbar} \cos(k_nL/2) \left( [ F_n(-L/2, t)-F_n(L/2, t)] + \sqrt{\frac{m}{\pi \hbar t}} [e^{\frac{i \pi \xi_n ^2(L/2, t)}{2}} - e^{\frac{i \pi \xi_n ^2(-L/2, t)}{2}}] \right)~,
\end{eqnarray}
which is zero for odd $n$. 
Note that one can find this, without doing any algebra. Wavefunction is an even (odd) function for odd (even) $n$ with respect to the point $x=L/2$, so its derivative is an odd (even) function for odd (even) $n$ with respect to that point. 

In the second case $\psi_0(x)=\frac{1}{(2\pi \sigma_0 ^2)^{1/4}} e^{-\frac{(x-x_0)^2}{4\sigma_0 ^2}}~\chi_{[0, L]}(x)$, in which $x_0$ is the center of the packet and $\sigma_0$ is its rms width; $\sigma_0 = \langle x^2 \rangle_0-\langle x \rangle^2_0$. After simultaneous removal of both walls, wavefunction is given by,
\begin{eqnarray}
\psi(x, t) &=& \frac{1}{(2\pi \sigma_0 ^2)^{1/4}} \sqrt{\frac{m}{2\pi i \hbar t}} \int_0^L dx^{\prime} e^{-\frac{(x^{\prime}-x_0)^2}{4\sigma_0 ^2} + \frac{im}{2\hbar t}(x-x^{\prime})^2}~.
\end{eqnarray}
\section{Bohmian trajectories} \label{Sec: BoTr}
In nonrelativistic Bohmian mechanics the world is described by point-like particles which follow trajectories determined by a law of motion. The evolution of the positions of these particles are guided by a wavefunction which itself evolves according to the Schr\"odinger equation. Given the initial position $x^{(0)} \equiv x(t=0)$ of a particle with the initial wavefunction $\psi_0(x)$, its subsequent trajectory $x(x^{(0)}, t)$ is uniquely determined by simultaneous integration of the time dependent Schr\"odinger equation, and the guidance equation $\frac{dx(t)}{dt}=v(x(t), t)$, in which $v=\frac{j}{\rho}$, where $j=\frac{\hbar}{m} \Im \left( {\psi^* \frac{\partial \psi}{\partial x}} \right)$ is the probability current density and $\rho=|\psi(x, t)|^2$ is the probability density. In the context of Bohmian mechanics arrival time distribution at a given location, say $x=X$ is given by \cite{Le-PhyslettA-1993,Le-book-1996,Le-PRA-1998},
\begin{eqnarray} \label{eq: arr_dis}
\Pi_X(\tau) &=& \frac{|j(X, \tau)|}{\int_0^{\infty} dt |j(X, t)|}~. 
\end{eqnarray} 
So, mean arrival time at observation point $x=X$ is determined by,
\begin{eqnarray} \label{eq: arr_mean}
\tau(X) &=& \int_0^{\infty} dt~t~\Pi_X(t)~.
\end{eqnarray} 
A general formulation for Bohmian arrival times was given in \cite{GrRh-JPA-2002} and a formula for the numerical calculation of such Bohmian arrival times in the case of 1D rigid inertial detectors (exactly the cases we have here) was presented in \cite{KrGrEm-JPA-2003}. The derived formula there doesn’t require the explicit calculation of the Bohmian trajectories and the resulting `cut-off current' can be considered to be a generalization of the arrival time probability density introduced by Leavens \cite{Le-PhyslettA-1993,Le-book-1996,Le-PRA-1998}.
According to Eq. (12) of \cite{KrGrEm-JPA-2003}, the probability density $\Pi_X(\tau)$ of the arrival time distribution for a point detector at $x=X$ takes the form,
\begin{eqnarray}\label{eq: ar_dis}
\Pi_X(\tau) &=& \left( \lim_{t \rightarrow \infty} P(t)\right)^{-1} j(X, \tau) \left[ \Theta\left( f_X(\tau)- {\text{max}} \{f_X (s)/0 \leq s \leq \tau \} \right) - \Theta\bigr(-f_X(\tau)- {\text{max}} \{-f_X (s)/0 \leq s \leq \tau \} \bigl) \right]~, 
\end{eqnarray}
in which $P$ is the detection probability, 
\begin{eqnarray} 
P(t) &=& {\text{max}} \{f_X (s)/0 \leq s \leq t \} + {\text{max}} \{-f_X (s)/0 \leq s \leq t \}~,
\end{eqnarray}
with 
\begin{eqnarray}
f_X(s) &=& \int_0^s j(X, t) dt~.
\end{eqnarray}
For the case of positive or negative $j(X, t)$ Eq. (\ref{eq: ar_dis}) reduces to Eq. (\ref{eq: arr_dis}).
%Our parameters for numerical calculations are such that $j(X, t)$ is positive at detector location $X$, so we use Eq. (\ref{eq: arr_mean}) to compute mean arrival time at detector location.
%
%
%
\section{Numerical results} \label{Sec: NuCa}
\begin{figure}
\centering
\includegraphics[width=10cm,angle=0]{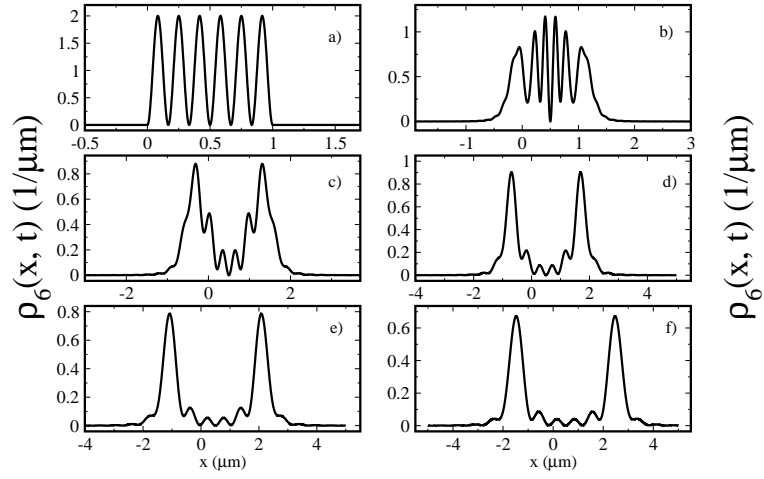}
\caption{Probability density versus distance $x (\mu$m) for state $n=6$ at times (a) $t=0$, (b) $t=0.03$ ms, (c) $t=0.06$ ms, (d) $t=0.09$ ms, 
(e) $t=0.12$ ms and (f) $t=0.15$ ms.}
\vspace*{.8cm}
\label{fig: den6di}
\end{figure}

\begin{figure}
\centering
\includegraphics[width=10cm,angle=0]{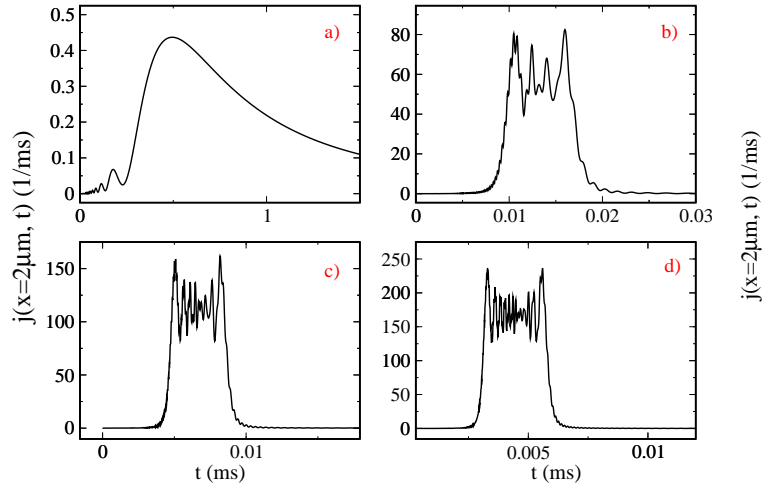}
\caption{Probability current density (1/ms) as a function of time $t$ (ms) at observation point $x=2~\mu$m for states (a) $n=1$, (b) $n=50$, (c) $n=100$ and (d) $n=150$.}
%\vspace*{.6cm}
\label{fig: curst}
\end{figure}

\begin{figure}
\centering
\includegraphics[width=10cm,angle=0]{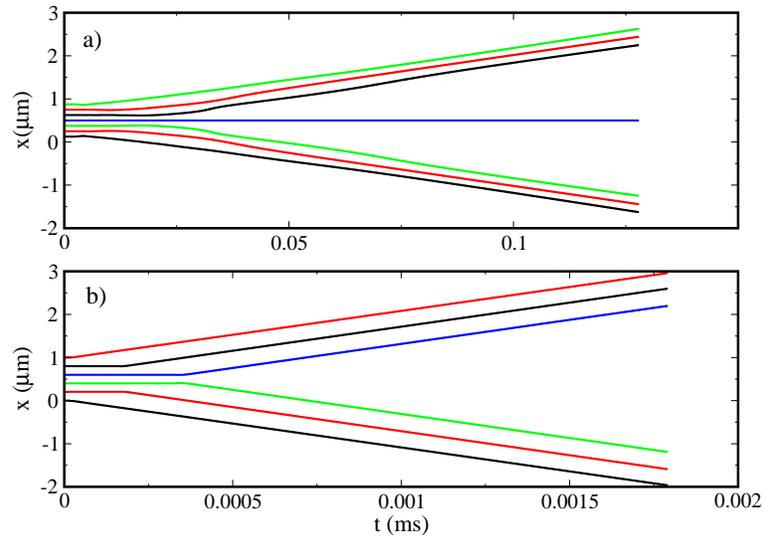}
\caption{A selection of Bohmian paths for states (a) $n=7$ and (b) $n=500$.}
%\vspace*{.6cm}
\label{fig: Bopa7_500}
\end{figure}

\begin{figure}
\centering
\includegraphics[width=10cm,angle=-90]{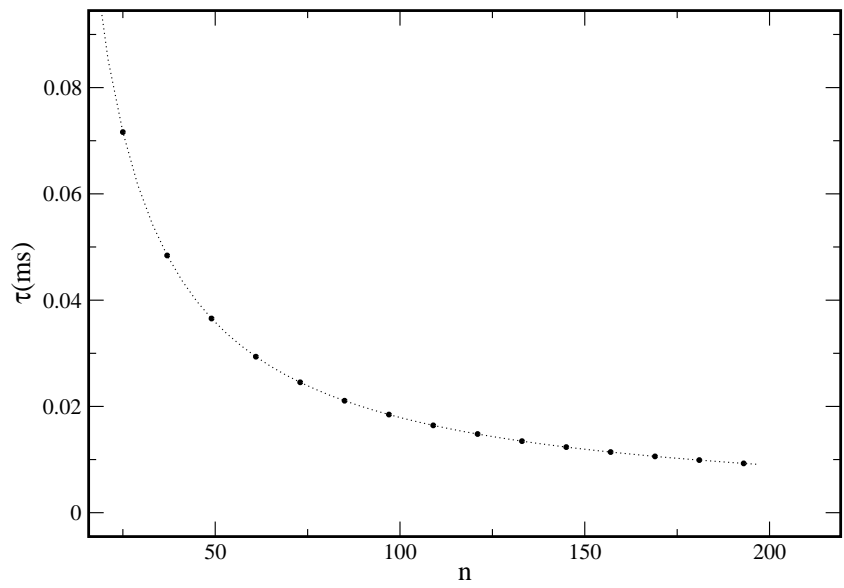}
\caption{Mean arrival time at detector position $x=2~\mu$m for different states.}
%\vspace*{.6cm}
\label{fig: mearst}
\end{figure}

For numerical calculation width of the well is chosen as $L=1~\mu$m. All of the calculations are presented for Rubidium atoms with mass $m=1.42 \times 10^{-25}~$kg. 
Fig. \ref{fig: den6di} shows probability density versus distance $x (\mu$m) for state $n=6$ at different times. At longer times there are two spatial packets placed about the box and moving apart. As pointed out by del Campo and Muga \cite{CaMu-EPL-2006} this takes place after the semiclassical time $t_n=mL^2/2n \pi \hbar$ as a result of the mapping of the underlying momentum distribution to the density profile expected asymptotically.
For our parameters $t_n=(0.214/n)$, hence, $t_7=0.031$ ms and $t_{500}=4.28 \times 10^{-4}$ ms. 
Fig. \ref{fig: curst} shows probability current density as a function of time at observation point $x=2~\mu$m outside the box after removal of the walls for various stationary states. 
From Eq.(\ref{eq: wf_n}) it is obvious that $x=L/2$ remains a node of the wavefunction for even $n$, i.e., $\psi _n(x=L/2, t)=0$ for even $n$. So because of a well-known property of Bohmian paths, Bohmian particle cannot be initially located at $x=L/2$ or even pass through this point for even values of $n$. As mentioned above, right after Eq. (\ref{eq: wf_der}), for odd $n$ distance-derivative of the wavefunction is zero at point $x=L/2$. Thus, current probability density and consequently Bohmian velocity is zero in this point all the time. Therefore, a Bohmian particle which is initially at $x^{(0)}=L/2$, will remain at rest. Because of noncrossing property of Bohmian path, particles with $x^{(0)}<L/2$ ($x^{(0)}>L/2$) will go backwards (forwards). See Fig. \ref{fig: Bopa7_500}. 
We have used Runge-Kutta method for the simultaneous integration of the time-dependent Sch\"{o}dinger equation and the guidance law to compute Bohmian trajectories.
For $n=7$, after $t= 0.03$ ms trajectories exhibit a bifurcation into two main branches while for $n=500$ this takes place after $t=0.0004$ ms. These values coincide very well with $t_7$ and $t_{500}$ in the above. 
%Congregation of trajectories happens at points where the effective force i.e., quantum $+$ classical force, is small.
Fig. \ref{fig: mearst} shows the mean arrival time, in the context of Bohmian mechanics, as a function of quantum number $n$ at the observation point $x=2~\mu$m. It is clear that $\tau$ decreases with $n$ as one expects, because by growing $n$, semiclassical velocity increases.

\begin{figure}
\centering
\includegraphics[width=10cm,angle=0]{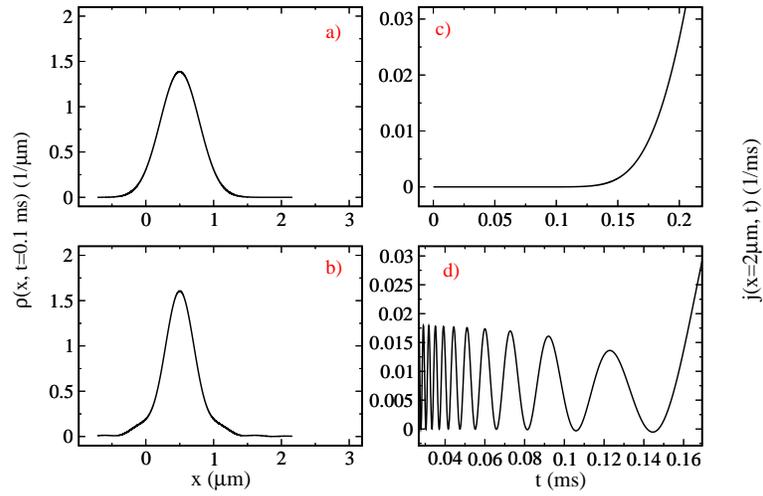}
\caption{Probability density $(1/\mu$m) versus distance $x (\mu$m) at time $t=0.1$ ms for (a) a free motionless Gaussian wave-packet and (c) a motionless truncated Gaussian wave-packet initially confined in a box. Probability current density (1/ms) versus time $t$ (ms) at observation point $x=2~\mu$m for (b) a free motionless Gaussian wave-packet and (d) a motionless truncated Gaussian wave-packet initially confined in a box.}
%\vspace*{.6cm}
\label{fig: dencurGadivati}
\end{figure}

\begin{figure}
\centering
\includegraphics[width=10cm,angle=-90]{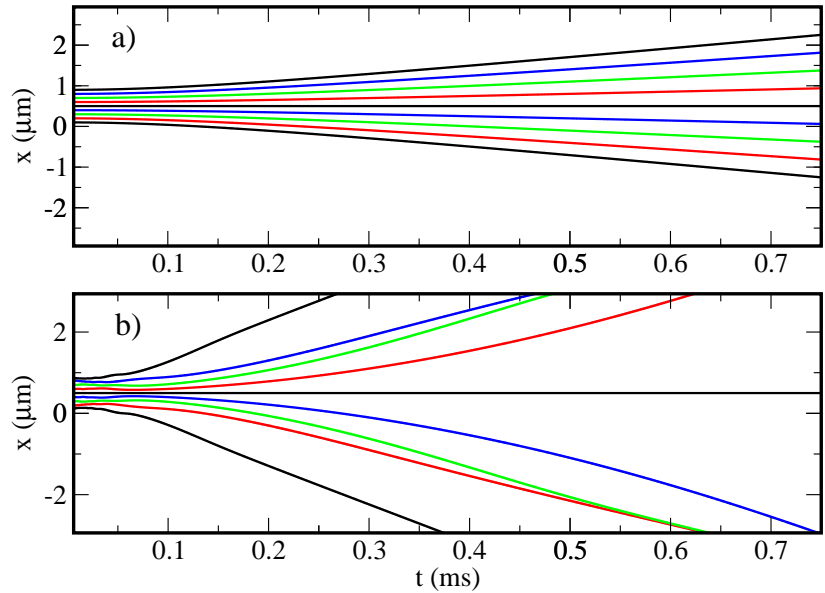}
\caption{A selection of Bohmian paths for (a) a free motionless Gaussian wave-packet and (b) a motionless truncated Gaussian wave-packet initially confined in a box.}
%\vspace*{.6cm}
\label{fig: BopaGa}
\end{figure}

In the case of a Gaussian packet, parameters of the packet are chosen as $x_0=0.5~\mu$m and $\sigma_0=0.25~\mu$m. It should be noted that with these parameters, initial Gaussian packet is not normalized to unity but to $0.954543$ (truncated Gaussian packet). If some one choses initial packet narrower than us, in such a way that it locates totally inside the well, then after removing the walls its evolution will be the same as a free Gaussian packet which is not desired here. To show the differences we consider a free Gaussian packet with the same parameters as well. Fig. \ref{fig: dencurGadivati} shows the probability density versus distance, $0.1$ ms after removal of the walls and the probability current density at observation point $x=2~\mu$m as a function of time for a free and a confined truncated Gaussian packet. In the confined case one sees some oscillations in the plot of current density which are absent in the free case. Finally, Fig \ref{fig: BopaGa} shows a selection of Bohmian paths for both cases. In the free case trajectories are determined by $x(t)=x_0+ (x^{(0)}-x_0)\sqrt{1+(\hbar t/2m\sigma_0^2)^2}$, where $x^{(0)}$ is the initial position of the particle\cite{Ho-book-1993}. From the figure it follows that the trajectory which starts at $x^{(0)}=x_0$ is the bifurcation trajectory in both cases and Bohmian velocity of a path in confined case is larger than the Bohmian velocity of the corresponding path in the free case.
\section{Summary and Conclusion} \label{CoRe}
The dynamics of particles, with various initial wavefunction, released from a box has been studied. 
Different boundary conditions, the absorbing wall and the reflecting one, is considered. Such studies suggest that the time-varying boundary conditions can give rise to interesting action-at-a-distance effects in quantum mechanics. Quantum temporal oscillations of matter waves released from a confinement region constitute the hallmark of the diffraction in time effect.
Mean arrival time at a observation point outside the box has been considered for various initial states. Moreover, Bohmian paths of the particle are computed. Our calculated mean arrival time may have no relevance to the experiment. Within Bohm's causal theory of quantum mechanics if we try to measure properties other than {\it position} (the only intrinsic property), we find that the result is affected by the process of interaction in a way that depends, not only on the total wave function, but also on the details of the initial conditions of both the particle and the apparatus. Our calculations can be verified if mean arrival times can be experimentally measured in such a way that the experiment does not perturb the unmeasured quantity.
\section{Acknowledgment}
We are very indebted to A. del Campo for helpful discussions, valuable suggestions and providing useful information.

\pagebreak
%********************************************************
%**********************************************************
%=========================================================
%=========================================================
%=========================================================
%=========================================================
%=========================================================
%=========================================================
\end{document}